\documentclass[journal,twoside,web]{ieeecolor}
\usepackage{tmi}
\usepackage{cite}
\usepackage{amsmath,amssymb,amsfonts}
\usepackage{algorithmic}
\usepackage{graphicx}
\usepackage{textcomp}
\usepackage{booktabs}
\usepackage{multirow}
\usepackage{graphicx}
\usepackage{subfigure}
\usepackage{array}
\usepackage{url}

\usepackage{float}

\def\BibTeX{{\rm B\kern-.05em{\sc i\kern-.025em b}\kern-.08em
    T\kern-.1667em\lower.7ex\hbox{E}\kern-.125emX}}
\begin{document}
\title{MouseGAN++: Unsupervised Disentanglement and Contrastive Representation for Multiple MRI Modalities Synthesis and Structural Segmentation of Mouse Brain}
\author{Ziqi Yu, Xiaoyang Han, Shengjie Zhang, Jianfeng Feng, Tingying Peng* and Xiao-Yong Zhang*
\thanks{This work was supported in part by National Natural Science Foundation of China under Grants 81873893, 82171903, 92043301; in part by Shanghai Municipal Science and Technology Major Project under Grant 2018SHZDZX01 and ZJLab. *Corresponding authors: Xiao-Yong Zhang: xiaoyong\_zhang@fudan.edu.cn and Tingying Peng: tingying.peng@helmholtz-muenchen.de}
\thanks{Z. Yu, X. Han, S. Zhang, J. Feng and X.-Y. Zhang are with Institute of Science and Technology for Brain-Inspired Intelligence, Fudan University, Shanghai, China. They are also with MOE Key Laboratory of Computational Neuroscience and Brain-Inspired Intelligence, and MOE Frontiers Center for Brain Science, Fudan University, Shanghai, China (e-mail: xiaoyong\_zhang@fudan.edu.cn)}
\thanks{T. Peng is with Helmholtz AI, Helmholtz zentrum Muenchen, Munich, Germany (e-mail: tingying.peng@helmholtz-muenchen.de).}
\thanks{\textbf{© 20xx IEEE. Personal use of this material is permitted. Permission from IEEE must be 
obtained for all other uses, in any current or future media, including 
reprinting/republishing this material for advertising or promotional purposes, creating new 
collective works, for resale or redistribution to servers or lists, or reuse of any copyrighted 
component of this work in other works}}}

\maketitle

\begin{abstract}

Segmenting the fine structure of the mouse brain on magnetic resonance (MR) images is critical for delineating morphological regions, analyzing brain function, and understanding their relationships. Compared to a single MRI modality, multimodal MRI data provide complementary tissue features that can be exploited by deep learning models, resulting in better segmentation results. However, multimodal mouse brain MRI data is often lacking, making automatic segmentation of mouse brain fine structure a very challenging task. To address this issue, it is necessary to fuse multimodal MRI data to produce distinguished contrasts in different brain structures. Hence, we propose a novel disentangled and contrastive GAN-based framework, named MouseGAN++, to synthesize multiple MR modalities from single ones in a structure-preserving manner, thus improving the segmentation performance by imputing missing modalities and multi-modality fusion. Our results demonstrate that the translation performance of our method outperforms the state-of-the-art methods. Using the subsequently learned modality-invariant information as well as the modality-translated images, MouseGAN++ can segment fine brain structures with averaged dice coefficients of 90.0\% (T2w) and 87.9\% (T1w), respectively, achieving around +10\% performance improvement compared to the state-of-the-art algorithms. Our results demonstrate that MouseGAN++, as a simultaneous image synthesis and segmentation method, can be used to fuse cross-modality information in an unpaired manner and yield more robust performance in the absence of multimodal data. We release our method as a mouse brain structural segmentation tool for free academic usage at https://github.com/yu02019.
\end{abstract}


\section{Introduction}
\label{sec:introduction}
\IEEEPARstart
{A}{ccurate} segmentation of brain structures on magnetic resonance (MR) images is crucial for delineating morphological structures, analyzing brain functions, and understanding their relationships. As one of the most important model organisms, the mouse plays an important role in neuroscience, drug discovery, and translational medicine. Since the mouse and human brains are evolutionarily conserved, the mouse brain has proven to be a powerful model for understanding human brain.

Currently, automatic segmentation methods of brain structures have been developed to segment human brains. However, when they are applied to MRI data of mouse brains, their performance remarkably suffers from the differences in image contrast, image size, and anatomy. Segmenting fine mouse brain structures on MRI data using automatic methods has remained a challenging task until now. The main reason is that accurate segmentation of fine brain structures usually requires multi-modality MRI with high-resolution, which provides much more complementary feature information than a single modality \cite{zhou2019reviewFusion}. For example, T1w images provide contrast differences between grey and white matter; T2w images are more sensitive to water-rich tissues; and quantitative susceptibility mapping (QSM) is suitable for differentiating deep brain tissues. These modalities described above provide complementary information for accurate segmentation of brain anatomical structures. However, such multi-modal mouse brain MRI data are often lacking because collecting such data takes too much scan time, which is impractical in most preclinical MR facilities. Therefore, mouse brain fine structure segmentation suffers from missing some MRI modalities in practice, but multi-modality fusion is expected to alleviate this dilemma.

Due to the restriction on acquiring a series of multi-modality images followed by segmentation, imputing the missing modality and decoupling semantic features has emerged as a crucial research area. In the computer vision field, image-to-image translation methods based on a generative adversarial network (GAN) have been demonstrated to be successful in image synthesis \cite{zhu2017cycleGAN, yi2017DualGAN, kim2017DiscoGAN}. Particularly, \cite{lee2020drit++} attempted to disentangle the modality-specific information and modality-invariant representations. In the medical image field, several existing cross-modality works have been applied to segment brain lesions \cite{yang2022D2-Net,chen2019robustRelated}. 
Whilst performing satisfactorily for human brain lesion segmentation tasks, these methods are not readily applicable to our mouse brain segmentation problem as they have several unsolved issues, such as patch-level context might not be appropriate for mouse brain regions due to their finer structures. Moreover, unlike lesions with distinguishable contrast on one specific MRI modality, there are often no significant contrast differences between different brain structures on single-modality MRI images. As a result, it is difficult to accurately delineate their anatomical boundaries using single-modal MRI data. To solve this problem, we aim to propose a synthesis-and-segmentation deep learning framework to first synthesize multi-modality MRI data, hence achieving more accurate segmentation of fine structures in the mouse brain via multimodal image fusion.

Specifically, we propose a disentangled and contrastive GAN-based framework, termed as MouseGAN++, which is a unified model that combines modality synthesis and structure segmentation. MouseGAN++ contains a modality translation module with two novel contrastive losses to project multi-modality image features into a shared latent content space that encodes modality-invariant brain structures and modality-specific attributes. Subsequently, the latent contents are combined with other modality-specific attributes to impute images of other modalities. Concretely, in MouseGAN++, the content contrastive loss is proposed to compel the network to avoid confusing structure-wise information during translation. 
We reuse attribute and content encoders during adversarial training to produce contrastive learning concurrently. The shared content space also facilitates decoder training in the segmentation module. Moreover, imputing modality with this model can enlarge the dataset, enabling the network to jointly learn modality-invariant semantic representations, thereby enhancing attention to multi-modality fusion.

This work significantly improves our previous model, MouseGAN, published in MICCAI conference \cite{yu2021MouseGAN}. Compared with MouseGAN and other related state-of-the-art (SOTA) work, MouseGAN++ has the following contributions:
\begin{itemize}
    \item MouseGAN++ uses an unsupervised disentanglement and contrastive framework that concurrently optimizes adversarial loss and contrastive loss without designing and pre-training on additional pretext tasks, thus eliminating the gap between pretext and translation tasks.
    \item MouseGAN++ disentangles MR images into attribute and content spaces, where the modality-agnostic content features can be flexibly recovered by various attributes or segmented by the decoder. Notably, two novel attribute and content contrastive losses introduced in MouseGAN++ further improve the disentanglement and the downstream segmentation performance. 
    \item We performed extensive experiments to evaluate both translation and segmentation tasks for the mouse brain, and provided the packaged MouseGAN++ as a pipeline that could further assist the mouse community.

\end{itemize}

\section{Related work}

\subsection{Mouse Brain Segmentation}

Brain structure segmentation in mouse MR images is a fundamental step in neuroimaging and preclinical studies. With the development of the active mouse research community, there have been many efforts to address this segmentation problem. To date, they can be mainly grouped into two categories based on their methodologies: atlas-based methods and deep learning methods. The atlas-based methods, which use image registration to propagate structural label information from a specific atlas volume to a native space corresponding to each subject volume, are primarily used in mouse brain segmentation pipelines \cite{cabezas2011reviewMouse,bai2012atlasMouse}. So far, only a few registration-based toolboxes, e.g., \cite{niedworok2016aMAP, bates2020Natverse}, can be used for brain segmentation. However, small and elongated regions are more susceptible to biases introduced by the registration procedure. Besides, the accessibility of suitable atlases and the complexity of making them further impede the spread of these atlas-based methods.

Recently, despite the successful application of deep learning in the field of segmentation, only a few attempts have been made to segment mouse brain structures, e.g., MU-Net \cite{de2021MU-Net}. However, due to the lack of prior knowledge of multi-modality contrasts, there is still a considerable performance gap when dealing with fine structure segmentation. In addition, how to disentangle these semantic representations in multimodal MRI data also presents challenges for the network design.

\subsection{Image Synthesis and Disentangled Representation}

Image synthesis techniques have made significant strides in recent years. Since modality synthesis can be deemed as an image-to-image translation task, GAN is considered the ideal model as it can be constrained on the conditional image, thus it has been widely used in medical images including MRI, CT, and PET \cite{chartsias2019disentangled, reaungamornrat2022multimodal-R1Q1,R2_wang20183d, R2_zhan2022d2fe}. Additionally, GAN brings strong disentanglement prior because of its hierarchical structure \cite{locatello2019ICML-best-paper}, which naturally serves our purpose.

CycleGAN is a seminal paper dealing with this task \cite{zhu2017cycleGAN}. Following the success of CycleGAN, recent efforts to exploit GAN mainly focus on the different construction strategies for the encoder and decoder, such as DiscoGAN \cite{kim2017DiscoGAN} and DualGAN \cite{yi2017DualGAN}. UNIT \cite{liu2017UNIT} proposes a shared latent space assumption for a better perceptual appearance. As most previous models consider a mapping between two domains, the scalability is limited when dealing with multiple-domain translation. Then MUNIT \cite{huang2018MUNIT}, DRIT++ \cite{lee2020drit++} and StarGAN-v2 \cite{choi2020starganV2} are proposed to address this limitation. Nevertheless, \cite{lee2020drit++, lee2018drit, huang2018MUNIT} assume the latent attribute spaces are under the constraint of the Gaussian prior, and \cite{choi2020starganV2} uses a Gaussian noise-into-style mapping network, which might lead to inadequate disentanglement. Thus, unlike the existing works, we explicitly impose a disentanglement prior to our MouseGAN++.

The core idea of disentangled representation learning approaches follows a prior that one model can learn to embed images into two spaces: a modality-specific attribute space and a modality-independent content space.
By assuming that a shared modality-independent content space that maintains structural information can be exploited for both modalities, previous work mainly concentrated on unsupervised domain adaptation. For instance, Chen et al. \cite{chen2020SIFA} proposes a method that conducts synergistic alignment of modalities from both image and feature perspectives. For fusing the complementary information from multimodal data, Chen et al. \cite{chen2019robustRelated} proposes a learning framework with feature disentanglement and gated feature fusion for robust segmentation and Yang et al. \cite{yang2022D2-Net} adopts contrastive loss for brain tumor segmentation, however, the paired training data they used is prohibitively expensive in clinics for the research. Recently, extensive efforts on GAN-based disentanglement methods have been explored. In \cite{yao2022AD-GAN-R5Q7-1, chartsias2018factorised, chartsias2019disentangled, chartsias2019multimodal, wagner2021structure-R4Q2-2, havaei2021conditional-R4Q2-3, wang2022sgdr-R4Q2-4, bercea2021FedDis-R4Q2-5}, they share the key idea of disentangling input images into modality-specific and modality-independent features. Since a majority of them introduce Gaussian priors into their latent attribute spaces, it might cause confusion in multi-modality feature distribution. In contrast to previous works, our method proposes contrastive learning prior into latent spaces by combining attribute features and shared content features with contrastive loss to facilitate learning various cross-modality mappings and improve segmentation performance.

\subsection{Contrastive Learning}
As a popular branch of self-supervised learning, the network guided by contrastive learning can learn hierarchical features from vast unlabeled datasets, which benefits the training of downstream tasks. Specifically, a contrastive loss is used to push positive pair representations to be similar and negative pair representations to be differentiated. SimCLR \cite{chen2020SimCLR} and MoCo \cite{he2020MoCo} are two representative methods that provide two training strategies with SOTA performance. Note that most contrastive learning methods focus on image classification, assuming that the instances in two individual images have distinct characteristics. However, when applied to brain MR images, the structural similarity patterns may confuse the network. For example,a recent work \cite{huang2022multi-R5Q7-2} utilizes the patch-wise contrastive losses for training. However, patch-level information fails to consider global context, as the same anatomical structure belonging to symmetrical left and right brain hemispheres should have similar features, so if we expect the network to distinguish them, it will lead to false negative results. Besides, the same structure in different modalities may present different features, which possibly results in a large number of false negative pairs. Several other works have also deployed contrastive learning in medical images successfully \cite{zeng2021positional, bercea2021FedDis-R4Q2-5}, however, the design of pretext tasks or the network architectures are decoupled from downstream applications. Inspired by SimCLR \cite{chen2020SimCLR}, we reuse two encoders in our model to seamlessly provide embedding features from each batch to conduct contrastive learning to guide both disentangled representation and the quality of generated images.

\begin{figure*}[t]
\centerline{\includegraphics[width=7.0in]{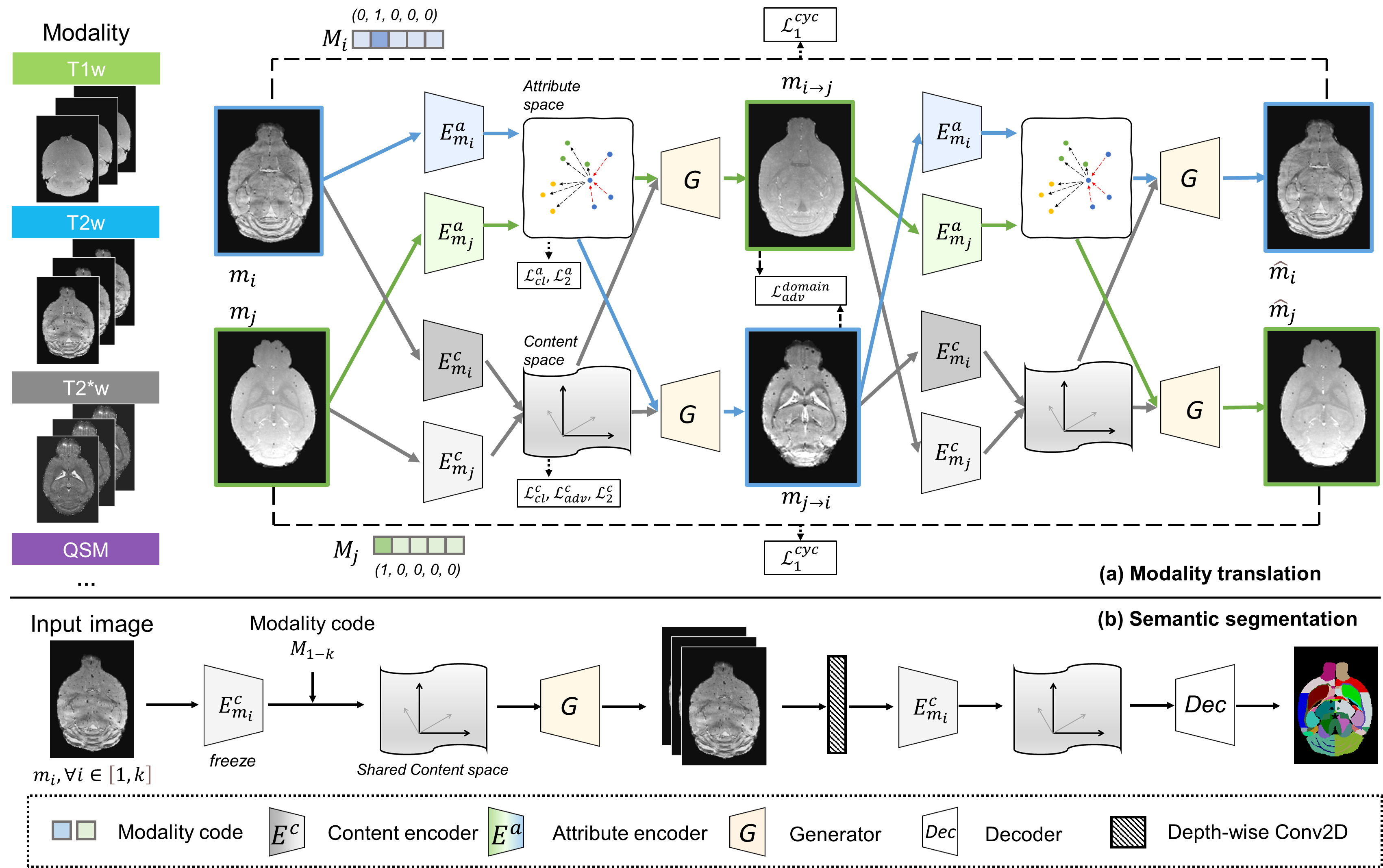}}

\caption{Schematic of MouseGAN++. Given an input which can be any of the five modalities (T1w, T2w, T2*w, QSM, and Mag), (a) we first train a modality translation module based on feature disentanglement and contrastive learning to synthesize all modalities. (b) Then, we employ this modality translation module as an auxiliary network in the subsequent segmentation pipeline by reusing the content encoder parameters and imputed missing modalities. We demonstrate that the learning of structural features shared between different modalities in a self-supervised manner can better characterize brain structures, thereby leading to better segmentation.}

\label{fig1}
\end{figure*}

\begin{figure*}[ht]
\centering
\includegraphics[width=0.75\textwidth]{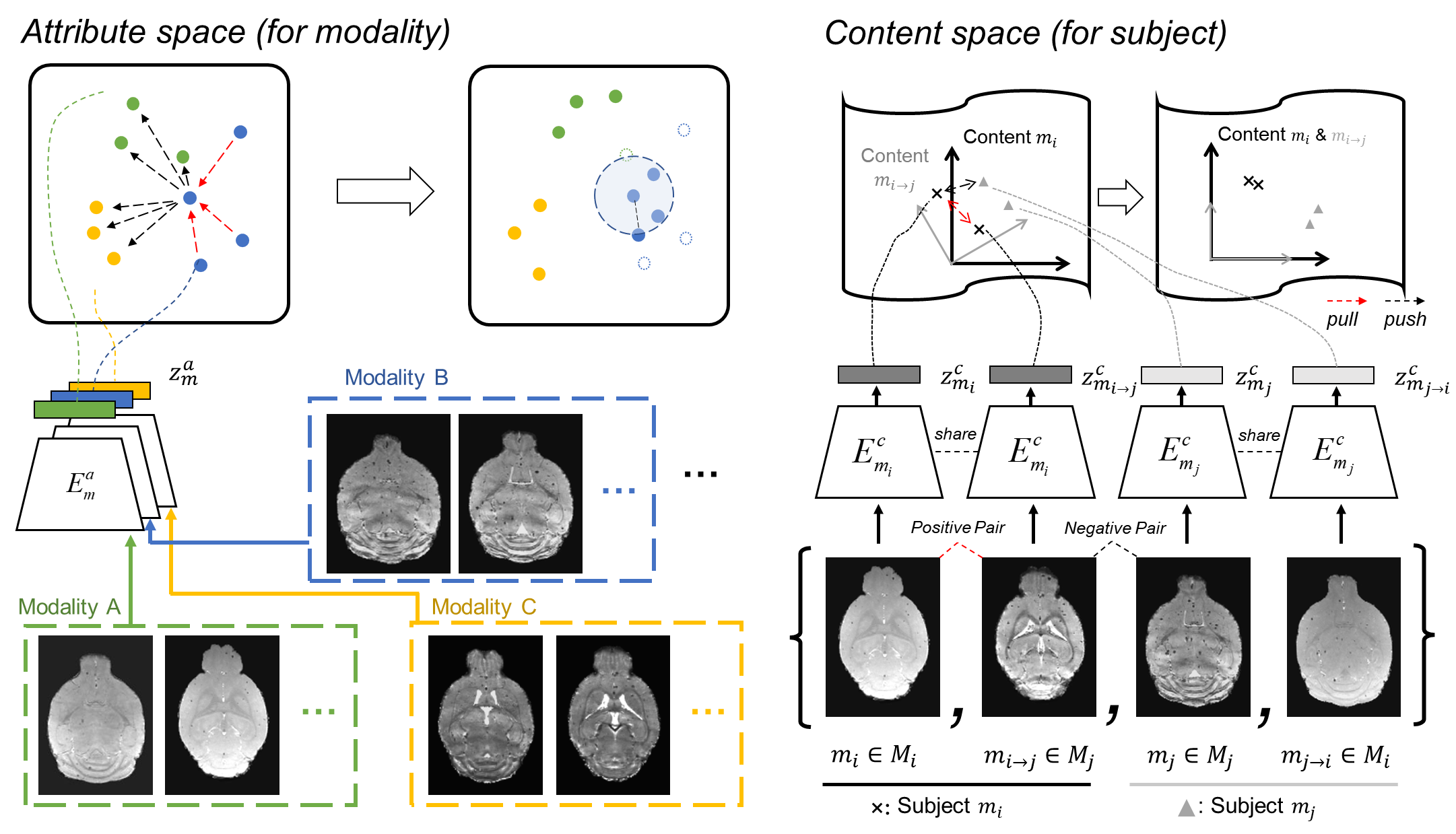}

\caption{Illustration of the attribute discrimination (left) and content discrimination (right) achieved by contrastive learning. In the attribute spaces, samples from the same modalities are defined as positive pairs and are driven close to each other, whilst samples from different modalities are negative pairs and are pushed away from each other. In the content space, samples from the same subject are defined as positive pairs and should have similar features in the embedding space, whilst samples from different subjects form negative pairs with distinct features.}
\label{fig2}
\end{figure*}

\section{Methods and concept formulation}
\subsection{Overview}

Fig. 1 depicts our proposed method for multi-modality image translation via unsupervised disentangled and contrastive representation, which improves the downstream segmentation model as it can leverage well-learned anatomical knowledge combined with imputed modalities which were originally missing.

The details of disentanglement learning are given in Section 3.B. To advance the disentangling operation and alleviate mode collapse during the training stage, we further propose two novel unsupervised contrastive learning strategies: one for attribute learning and the other for content learning as add-on inductive priors detailed in Section 3.C and 3.D, respectively. Together, they complement disentanglement learning as the attribute contrastive learning forces the modality-specific features to be separable in latent attribute space, whilst the content contrastive learning ensures subject-specific features are aligned in content space. The mathematical formulation of each loss function in the modality transfer module is given in Section 3.E. After successful feature disentanglement by the above-mentioned modality transfer module, we could utilize the learned content encoder as the encoder of our segmentation model, as stated in Section 3.F. As the building blocks are adopted from our previous MouseGAN \cite{yu2021MouseGAN}, which uses a modified version of DRIT++ \cite{lee2020drit++} as the translation module, the content and attribute encoders' architectures remain the same as \cite{yu2021MouseGAN} and then be reused seamlessly for contrastive losses.

\subsection{Basic Disentanglement Building Blocks}
Based on the consensus that unsupervised disentanglement is impossible unless appropriate inductive biases are imposed \cite{locatello2019ICML-best-paper}, we introduce our basic assumption here that anatomical semantic information should be included in content factors, whereas style factors only differ in image appearance. We believe this assumption should fit well with the fundamental nature of multi-modality MR images. Thus, in this section, we describe how we decouple multi-modality images into modality-specific attribute space and subject-specific (and modality-independent) content simultaneously, and then further introduce the derivation of contrastive learning priors in the subsequent sections.

Denoting $k$ as the number of modalities in the dataset, $ \{ M_i \}_{i=1 - k}$ be the descriptor for each modality, $(m_i ,m_j) \in M $ be the two randomly-sampled images from two exemplary modalities, respectively, and their corresponding modality codes in one-hot format $( z_{m_i}^d,z_{m_j}^d )$, where $z^d \subset R^k$. As shown in Fig. 3, the model consists of modality-specific content encoders $ \{ E_{m_i}^c,E_{m_j}^c \}$, attribute encoders $ \{ E_{m_i}^a,E_{m_j}^a \}$, generators $ \{ G_{m_i}^c,G_{m_j}^c \}$, and modality discriminators $ \{ D_{m_i},D_{m_j} \}$ as well as a content discriminator $D^c$. In the encoding path, for given input images $(m_i ,m_j)$ , we obtain the disentangled modality-independent content features $z_{m_i}^c = E_{m_i}^c(m_i)$, $z_{m_j}^c = E_{m_j}^c(m_j)$ in the shared content space $ \{ \Omega^{c}_i \}_{i=1 - k}$ via the content encoders and simultaneously, the modality-specific attribute codes $z_{m_i}^a=E_{m_i}^a(m_i)$ and $z_{m_j}^a=E_{m_j}^a(m_j)$ in the latent attribute space $ \{ \Omega^{a}_i \} _{i=1 - k}$ via the attribute encoders.
The first forward image style translation cycle is achieved by a generator by combining the content feature from $m_i$ as well as the swapped attribute feature from $m_j$, i.e., $m_{i\rightarrow j}{=G}_{m_j} (z_{m_i}^c,z_{m_j}^a,z_{m_j}^d )$. The disentanglement of modality-independent content features $z_{m_i}^c$ is desirable for us as it well conserves the structural properties which could be crucial for  medical image synthesis and downstream applications, albeit the generated image $m_{i \rightarrow j}$ have a similar high-level style representation like $m_j$. Similarly, we can generate $m_{j\rightarrow i}{=G}_{m_i} (z_{m_j}^c,z_{m_i}^a,z_{m_i}^d )$ by exchanging its attribute and modality codes $z_{m_j}^a$, $z_{m_j}^d$ for $z_{m_i}^a$, $z_{m_i}^d$ while preserving its own content feature $z_{m_j}^c$.

\subsection{Modality-Specific Attribute Discrimination}

To promote learning a better disentangled feature representation, we introduce a task-specific prior here, as our desirable attribute embedding space should satisfy the following requirements:
\begin{itemize}
    \item The attribute features of the same modality, even those from different subjects, are close to each other in latent attribute space.
    \item The attribute features of different modalities, even those from the same subject, are pushed away.
\end{itemize}
	
These requirements are in accordance with the key idea from contrastive learning theory \cite{tian2020makes,khosla2020supervisedNIPS}, which learns feature representation for positive pairs to be similar while pushing features from negative pairs apart. To this end, inspired by the work of \cite{he2020momentum,chen2020SimCLR, ye2019unsupervised, khosla2020SuperCL}, we introduce contrast learning to guide the attribute encoder to achieve a more effective and discriminative modality-specific attribute embedding. Specifically, considering we have two input images $m_i$ and $m_j$ in one batch. Note that $i$ could be equal to $j$ here, when the two randomly sampled images are coincidentally from the same modality. After feeding them into the attribute encoder $E_{m_i}^a$  and $E_{m_j}^a$, we have embedding attribute features $ z_{ m_i}^a = E_{m_i}^a (m_i)$ and $ z_{ m_j}^a = E_{m_j}^a (m_j)$, as shown in Fig. 2. Unlike the most commonly used contrastive learning technology that employs instance discrimination, e.g., regarding each image as a separate individual, we extend the contrastive learning concept to attribute (modality) discrimination here. More specifically, the positive pairs contributed to $ \Omega_{a}^+ = \{ {m_i, m_{j \rightarrow i} | \forall i,j } \} $ are defined by the images that share the same modality code $z_{m_i}^d$, which could be either originated from or translated to the same modality, whereas the the negative ones $ \{ {m_j, m_{i \rightarrow j} | \forall i,j \, \text{and} \, i \neq j } \} $ from different modalities. We optimize the attribute space by minimizing the contrastive loss, which could be formulated as follows:
\begin{equation}
{\cal L}_{cl}^{a} = \sum_{i=1}^N \frac{-1}{| \Omega_{a}^+ |} \sum_{m^+ \in \Omega_{a}^+ } \log \frac{{\exp \left( { CL^+ /{\tau _{a}}} \right)}}{{ \sum_{m_a \in \Omega_a (m_a)} {\exp \left( {  CL /{\tau _{a}}} \right)} }},     \label{eq1}
\end{equation}
where $ CL^ + = sim ( z_{ m_i}^a , z_{m^+}^a ) $, $ CL =  sim ( z_{ m_i}^a ,  z_{m_a}^a  ) $ and $ \Omega_a (m_a) = \Omega_a \backslash \{ m_i \} $ . $\tau_{a}$ is the temperature scaling parameter. $sim( \cdot , \cdot)$ is pairwise similarity function which calculates the similarity of two vectors in the attribute space and is defined by cosine distance:
\begin{equation}
sim(x,y) = \frac{{x \cdot y^\top }}{{\left\| x \right\| + \left\| y \right\|}}.     \label{eq2}
\end{equation}

\begin{figure}[h]
\centering
\includegraphics[width=0.48\textwidth]{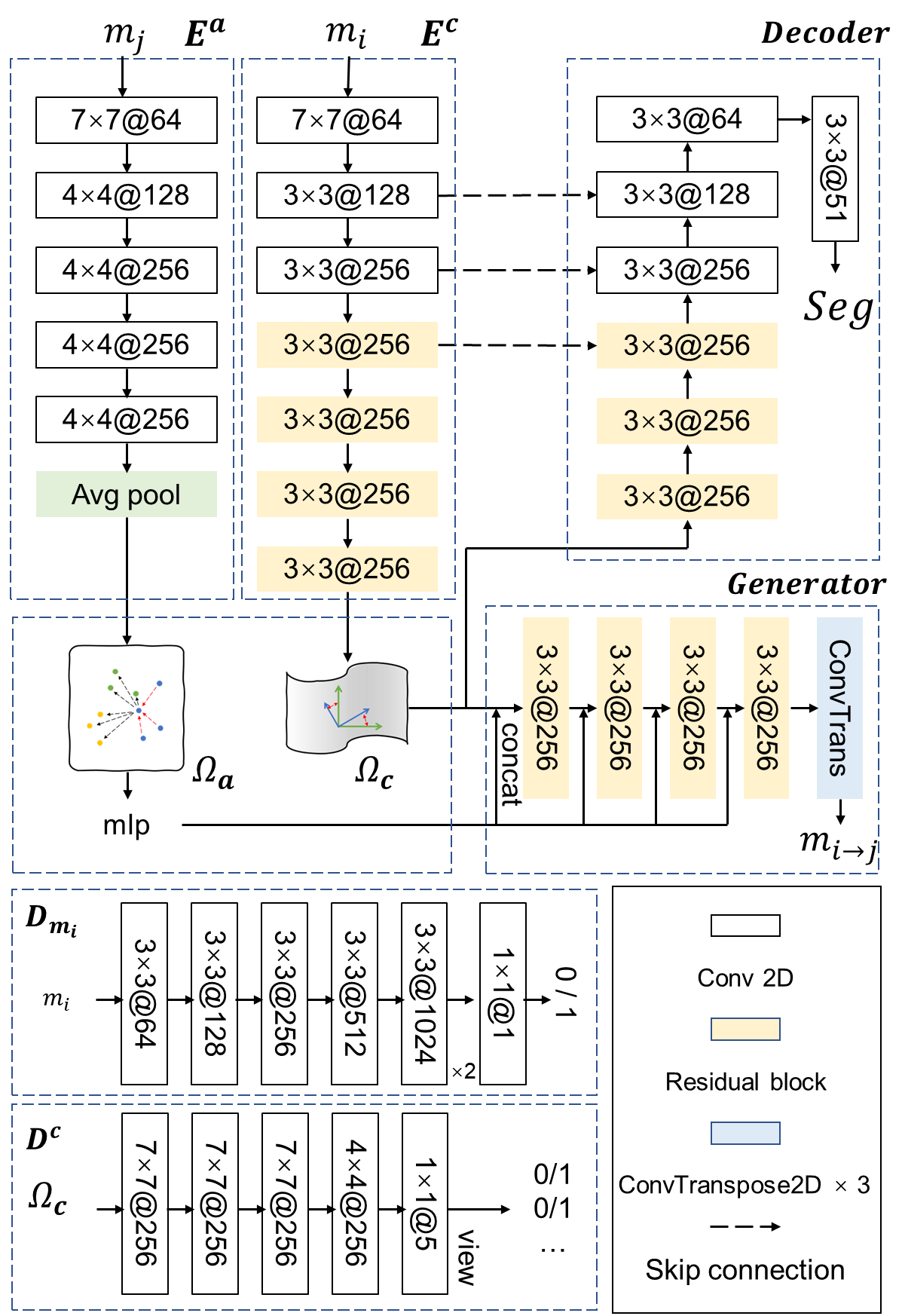}

\caption{The network architecture of attribute ($ E^a $), content ($ E^c $) encoder, decoder, generator and discriminator ($ D_{m_i} \text{/} D_{m_j} $ and $D^c$) in our framework.}
\label{fig3}
\end{figure}

\subsection{Instance-Specific Content Discrimination}

In medical image synthesis, one of the grimmest issues to contend with is how to retain the anatomical structure information fidelity during translation. Many recent image synthesis methods mainly focus on global visual similarity or the rationality of perception. These may work well for natural images but are difficult to transfer to medical images in practice. Here, we propose an inductive prior that the content features $z_{m_i}^c$ are modality-agnostic and preserve only structural information, which should be subject-specific. Hence, we formulate two criteria to constrain the content space:

\begin{itemize}
    \item For a specific subject, the content features of input image $m_i$ and translated images $m_{i \rightarrow j}$ are close to each other in latent content space since they share the same intrinsic structures.
    \item For different subjects, the content features are pushed away, regardless of their image modalities.
\end{itemize}
	
Concretely, as shown in Fig. 2, given the images $m^1_i$ and $m^2_i$ from two subjects, the embedding content features $z_{m^1_i}^c = E^c (m^1_i)$ and $z_{m^2_i}^c = E^c (m^2_i)$ are generated via content encoders, whereas their translated content features are denoted by $z_{m^1_{i \rightarrow j}}^c$ and $z_{m^2_{i \rightarrow j}}^c$, respectively. Similarly as attribute discrimination in above section, we design the content discrimination pretext task here, which denotes $m^1_i$ and $m^1_{i \rightarrow j}$ as a positive pair $\in \Omega_c^+ $ (same subject, different modalities) but $m^1_i$ and $m^2_{j \rightarrow i}$ as a negative pair (different subject, same modalities). Mathematically, we have:
\begin{equation}
{\cal L}_{cl}^{c} = \sum_{i=1}^N \frac{-1}{| \Omega^+_c |} \sum_{m^{1+} \in \Omega^+_c } \log \frac{{\exp \left( { CL^+ /{\tau _{c}}} \right)}}{{ \sum_{m_c^1 \in \Omega_c (m_c^1)} {\exp \left( {  CL /{\tau _{c}}} \right)} }}.
\label{eq}
\end{equation}
where $ CL^ + = sim ( z_{ m_i^1}^c , z_{m^+}^c ) $ and $ CL =  sim ( z_{ m_i^1}^c ,  z_{m_c}^c  ) $ and $  \Omega_c (m_c^1) = \Omega_c \backslash \{ m_i^1 \} $. $\tau_{c}$ is the temperature scaling parameter. In our experiments, we set temperature of $\tau_{c}$, $\tau_{a}=0.1$.

\subsection{The Overall Loss Function of Modality Translation}
As bidirectional reconstruction loss further encourages disentanglement by synergistically updating the gradient propagation among blocks, we reconstruct synthetic images $(m_{i \rightarrow j}, m_{j \rightarrow i} )$ back via a second modality transfer step and enforce $(\hat{m}_i,\hat{m}_j )$ to match the real input image $(m_i,m_j )$ with cross-cycle consistency loss:
\begin{equation}
\begin{aligned}
\hspace {-0pc}
{\cal L}_1^{cyc} ( E_{m}^c,E_{m}^a,{G_{m}}  ) = 
\Vert \hat{m}_i - m_i  {\Vert _1}
+ \Vert \hat{m}_j - m_j   {\Vert _1},
\end{aligned}
\end{equation}

In addition, to compel the task of unsupervised translation during training, we use a self-reconstruction loss $ {\cal L}_1^{self-recon}$ to constrain encoders and decoders for the generation of high quality translations, and can be formulated as follows:
\begin{equation}
\begin{aligned}
& \hspace {-0.5pc}  {\cal L}_1^{self-recon} \left( {E_{m}^c,E_{m}^a,{G_{m}}} \right)  = \\& \Vert {G_{m_i}} ( z_{m_i}^c, z_{m_i}^a ) - {m_i}{\Vert _1}  +  \Vert {G_{m_j}} ( z_{m_j}^c, z_{m_j}^a ) - {m_j}{\Vert _1},
\end{aligned}
\end{equation}

The latent space reconstruction loss $ {\cal L}_1^{latent} $ is adopted from \cite{zhu2017LatentLoss}. The $ {\cal L}_{cls}^{domain} $ is used to classify the modality of images \cite{lee2018drit}. To enforce the translated images to be as real as possible, following the successful practice of \cite{zhu2017cycleGAN,yu2021MouseGAN,lee2020drit++}, the discriminators $ \{ D^c, D_{m_i}, D_{m_j}\} $ and corresponding losses $ \{{\cal L}_{adv}^{cont}, {\cal L}_{adv}^{domain} \} $ are implemented for adversarial learning: the former discriminator enforces that one cannot infer image modality from content codes only, whilst the later one guarantees the synthesized images has the modality-specific features as the real ones. We formulate these losses as follows:
\begin{equation}  
\begin{aligned}
& \hspace {-2pc} 
{\mathop {\min }\limits_{\left( {E^{c},{G}} \right)} \mathop {\max }\limits_{D^c} {\cal L}_{adv}^{cont}} =
\\& \hspace {-0.7pc}  \mathbb {E}_{m_i}[ \frac{1}{2}  \text {log}D^c ( z^c_{m_i} ) +  \frac{1}{2}  \text {log}(1 - D^c ( z^c_{m_i} )) ] + \\& \hspace {-0.7pc} \mathbb {E}_{m_j}[ \frac{1}{2}  \text {log}D^c ( z^c_{m_j} ) +  \frac{1}{2}  \text {log}(1 - D^c ( z^c_{m_j} )) ],
\end{aligned}
\end{equation}
\begin{equation}
\begin{aligned}
& \hspace {-1pc} 
{\mathop {\min }\limits_{\left( {E^{c},E^{a},{G}} \right)} \mathop {\max }\limits_{D} {\cal L}_{adv}^{domain}} =
\\& \mathbb {E}_{m_i}[ \text {log}D_{m_i} (m_i) ] +  \mathbb {E}_{m_{j\rightarrow i}}[ \text {log}(1- D_{m_i} (m_{j\rightarrow i})+ \\& \mathbb {E}_{m_j}[ \text {log}D_{m_j} (m_j) ) ]   + \mathbb {E}_{m_{i\rightarrow j}}[ \text {log}(1- D_{m_j} (m_{i\rightarrow j})],
\end{aligned}
\end{equation}

Besides, we impose L2 regularization $ {\cal L}_2^{c, a} $ to make latent space compact. The overall objective loss function of the translation module is:
\begin{equation}
\begin{aligned}
& {\cal L}_{trans} = {\cal L}_{adv}^{domain} + {\cal L}_{adv}^{cont} + {\lambda _1}{{\cal L}_1^{self-recon}} + {\lambda _2}{ {\cal L}_1^{cyc}} + \\&
 {\lambda _3}{ {\cal L}_1^{latent}} + {\lambda _4}{ {\cal L}_{2}^{c, a}} + {\lambda _5}{ {\cal L}_{cls}^{domain}} + {\lambda _6}{ {\cal L}_{cl}^{a}} + {\lambda _7}{ {\cal L}_{cl}^{c}} .
\end{aligned}
\end{equation}

\subsection{Semantic Segmentation Model}
The pretrained modality translation model above serves as an auxiliary network for image segmentation by imputing missing modality in the input, since it is too expensive and time-consuming to collect all modality data for every mouse. After modality translation, an original image and its synthesized ones are fed to a depth-wise convolution before an encoder to preprocess the context modality-by-modality (Fig. 1b). We reuse the architecture and the parameters of the content encoder obtained in the modality translation training stage as the encoder of the segmentation module and update the decoder first. This is motivated by the fact that the content encoder in the modality translation model has already learned modality-independent anatomical features in a self-supervised manner and could extract and distill these representative features in a shared latent content space, leading to a better segmentation of anatomical structures for different modalities. Moreover, in contrast to a typical segmentation task, which is only trained by labelled data in a supervised fashion, our segmentation model also leverages information from unlabeled data via the unsupervised modality translation model. Our segmentation loss is defined as:
\begin{equation}
\begin{aligned}
{\cal L}_{seg} = - \sum _{k=1}^K (y_i)_j^k \log (Decoder(E^c(m_i)))_j^k .
\end{aligned}
\end{equation}
where $ y_i $ is the segmenation ground truth of $ m_i $, $ Decoder(\cdot)_j^k $ denote the probability prediction of voxel $j$ for class $k$.

\section{Experimental Setup}
\subsection{Dataset Acquisition}
\subsubsection{Multi-Modality Dataset}

The multi-modality structure images of the mouse brain were generated by a 3D-mGRE sequence acquired with a 11.7T MR scanner, including T2w imaging, T1w imaging, T2*w imaging, quantitative susceptibility mapping (QSM), and Magnitude MR images (Mag). 3D-mGRE was acquired with the following parameters: $TR/TE/\Delta TE = 100/2/2 \ ms$, echo number = 12, flip angle = $15^{\circ}$, and resolution = $0.07 \times 0.07 \times 0.07 \ mm^3$. The multi-modality structure images covered 75 mouse brains with 56,970 MR slices. All images were preprocessed by skull stripping and bias field correction. The ground truth of 50 brain structures was generated via an atlas-based method and then manually corrected by a brain anatomy expert.

\subsubsection{MRM NeAt Dataset}
To further verify our proposed method, we also tested it on external data from the MRM NeAt dataset \cite{ma2008MRM-dataset}, which includes 10 T2w MR images acquired with a 16.7T MR scanner. Each scan was manually annotated into 37 structures. All images had the resolution = $0.10 \times 0.10 \times 0.10 \ mm^3$, covered 960 slices, and were preprocessed by skull stripping, denoising, and bias field correction. Since this dataset included mono-modality, we only executed segmentation tasks on this dataset for comparison.

\subsection{Implementation Details}
\subsubsection{Multi-Modality Datset}

The architecture of MouseGAN++ is presented in Fig. 3. To conduct 5-fold cross-validation, we randomly split $80\%$ scans (60 subjects) as training sets and $20\%$ scans (15 subjects) as testing sets at subject-level in each modality and further utilize unpaired data in experiments for both translation module and segmentation module training. In the preprocessing process, each slice was resized to $ 256 \times 256 $ matrix size and the intensity distribution was normalized into zero mean and unit variance. Data augmentation was also introduced to alleviate overfitting, including random cropping to $ 216 \times 216 $ matrix size and random flip. In addition, to train the translation module, we used one-hot modality codes to indicate each modality of input images in the training stage. In the test stage, the input single-modality images were encoded into shared content space and generated to all modalities with different one-hot modality codes. Empirically, we set $ \lambda_1 , \lambda_2 , \lambda_3=10 $, $ \lambda_4=0.01 $, $ \lambda_5, \lambda_6, \lambda_7=1 $.  

For segmentation model training, we first froze the parameters of the shared content encoder and then trained only the decoder and depth-wise convolution filter. The images are resized to $ 224 \times 160 $. For all training procedures, we used the Adam optimizer with a learning rate of 0.0001, and set the size of the batch to 16. Finally, we finetuned the segmentation model with a learning rate of 0.00001 to update and refine all parameters, including the content encoder.
The proposed framework was deployed in the Pytorch library and trained on a NVIDIA Tesla V100 GPU (32 GB memory) with 400 and 100 epochs for the translation and segmentation modules, taking 40 and 8 hours on average, respectively.

\subsubsection{MRM NeAt Dataset}
On this external dataset, we executed only the segmentation module of MouseGAN++ to investigate how the disentanglement representation pretrained on multiple modalities facilitates downstream segmentation tasks. We conducted the 5-fold cross validation study at the subject-level, using 8 subjects for training and the remaining 2 subjects for testing in each equal-size fold. The parameter setting and training steps were kept in line with the multi-modality dataset other than the number of output channels.

\subsection{Evaluation Metrics}
These metrics were used to evaluate the quality of translation images, Learned Perceptual Image Patch Similarity (LPIPS) \cite{zhang2018LPIPS}, Visual Information Fidelity (VIF) \cite{sheikh2006VIF}, Peak Signal Noise Rate (PSNR), Structural Similarity Index Measure (SSIM), and Multi-Scale Structural Similarity (MS-SSIM). Lower LPIPS, higher VIF, PSNR, SSIM, and MS-SSIM refer to a better translation. The dice coefficient and average surface distance (ASD) were used for evaluating segmentation performance at subject-level in 3D volumes for both multi-modality dataset and MRM NeAt dataset. All the results were presented with mean$\pm$std to exhibit both the mean performance and cross-subject variance.

\subsection{Comparison to SOTA Methods}
We first compared the translation module of MouseGAN++ with the following state-of-the-art (SOTA) image translation methods: CycleGAN \cite{zhu2017cycleGAN}, UNIT \cite{liu2017UNIT}, MUNIT \cite{huang2018MUNIT}, StarGAN-v2 \cite{choi2020starganV2} and our prior MouseGAN \cite{yu2021MouseGAN}. As for segmentation, we further compared our method with two atlas-based methods, aMAP \cite{niedworok2016aMAP} and Natverse \cite{bates2020Natverse}; two direct segmentation methods: U-Net \cite{ronneberger2015U-Net} and MU-Net \cite{de2021MU-Net}; one disentanglement method: $D^2$-Net \cite{yang2022D2-Net}; in addition to four synthesis-and-segmentation methods: CycleGAN \cite{zhu2017cycleGAN}, SynSeg\cite{huo2018synseg}, UNIT \cite{liu2017UNIT}, MUNIT \cite{huang2018MUNIT}. For SOTA synthesis-and-segmentation methods, we trained a segmentation network on both real images in the original modality $ M_i $ and the synthetic ones $ M_{j \rightarrow i } $ in the transferred modalities, thus the network training can be benefited from the expansion of training data both in terms of size and modality. For SynSeg revised from \cite{huo2018synseg}, we conducted translation and segmentation simultaneously, not separately like in CycleGAN. As for MouseGAN and MouseGAN++ (the backbone, U-Net, is identical to that utilized in comparison methods), we utilized the translation module in our framework for imputing modalities, allowing the fused multi-modality semantic information to directly pass into the segmentation network. All comparison methods are either directly taken from original implementation from the corresponding github repository (if the codes are released) or implemented followed the original paper.
The impact of each module in MouseGAN++ is detailed in the ablation study. Additionally, we directly deployed U-Net \cite{ronneberger2015U-Net} as a baseline with only real images input for training. Since synthesis-and-segmentation methods require multiple modalities for image synthesis training, they fail to apply to the MRM NeAt dataset, which contains only one modality (T2w). Thus, we compared U-Net, the backbone of synthesis-and-segmentation methods, as well as other SOTA methods. Then, using several metrics, we evaluated the segmentation performance of these methods on several critical brain structures, including hippocampus, superior colliculus, striatum, and thalamus.

\begin{table}[h]
\centering
\caption{Translation performance comparison with different methods from T1w to T2w and vice versa}
\includegraphics[width=0.48\textwidth]{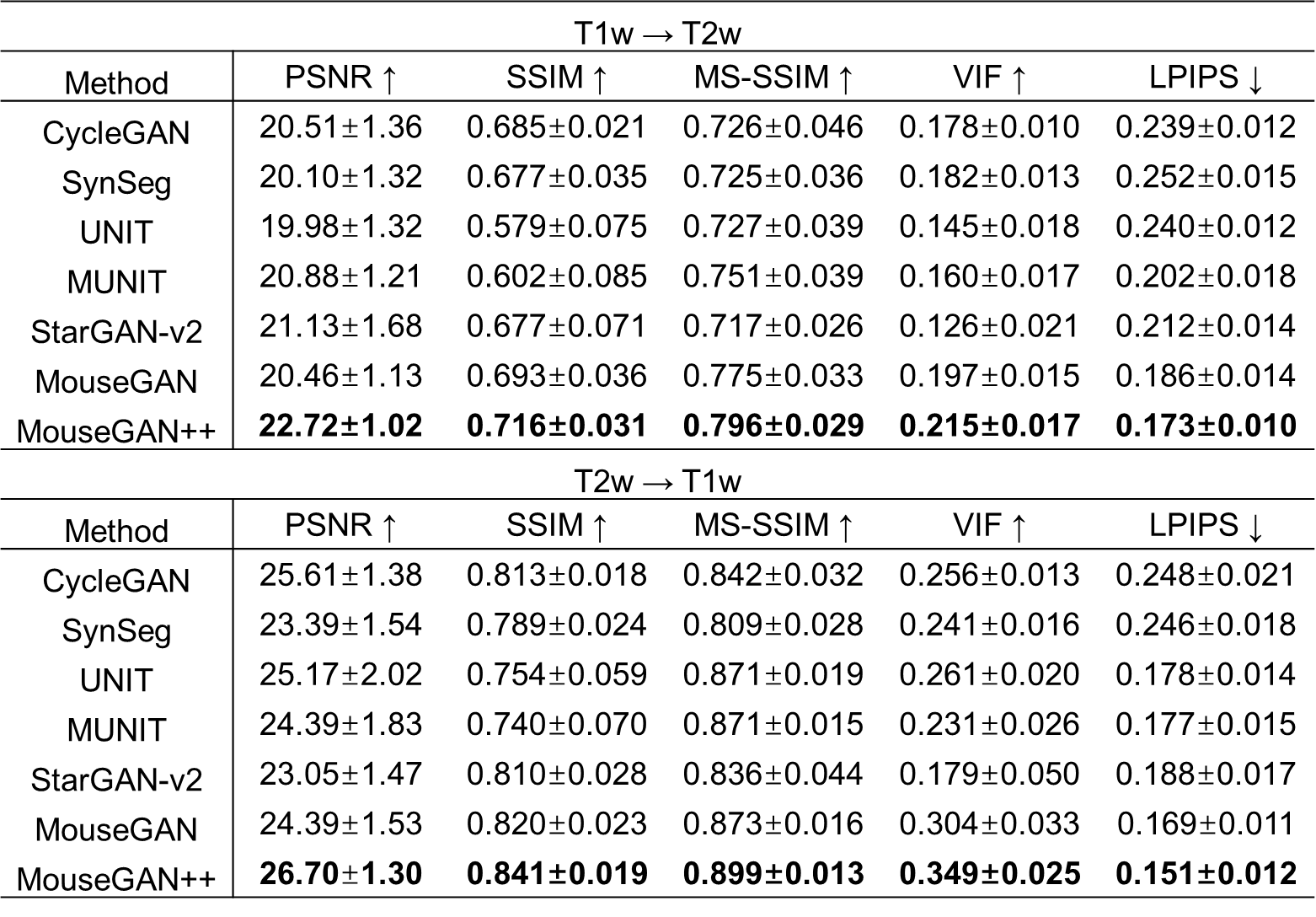}
\end{table}

\section{Results and Discussion}

\begin{figure*}[!h]
\centering
\includegraphics[width=0.95\textwidth]{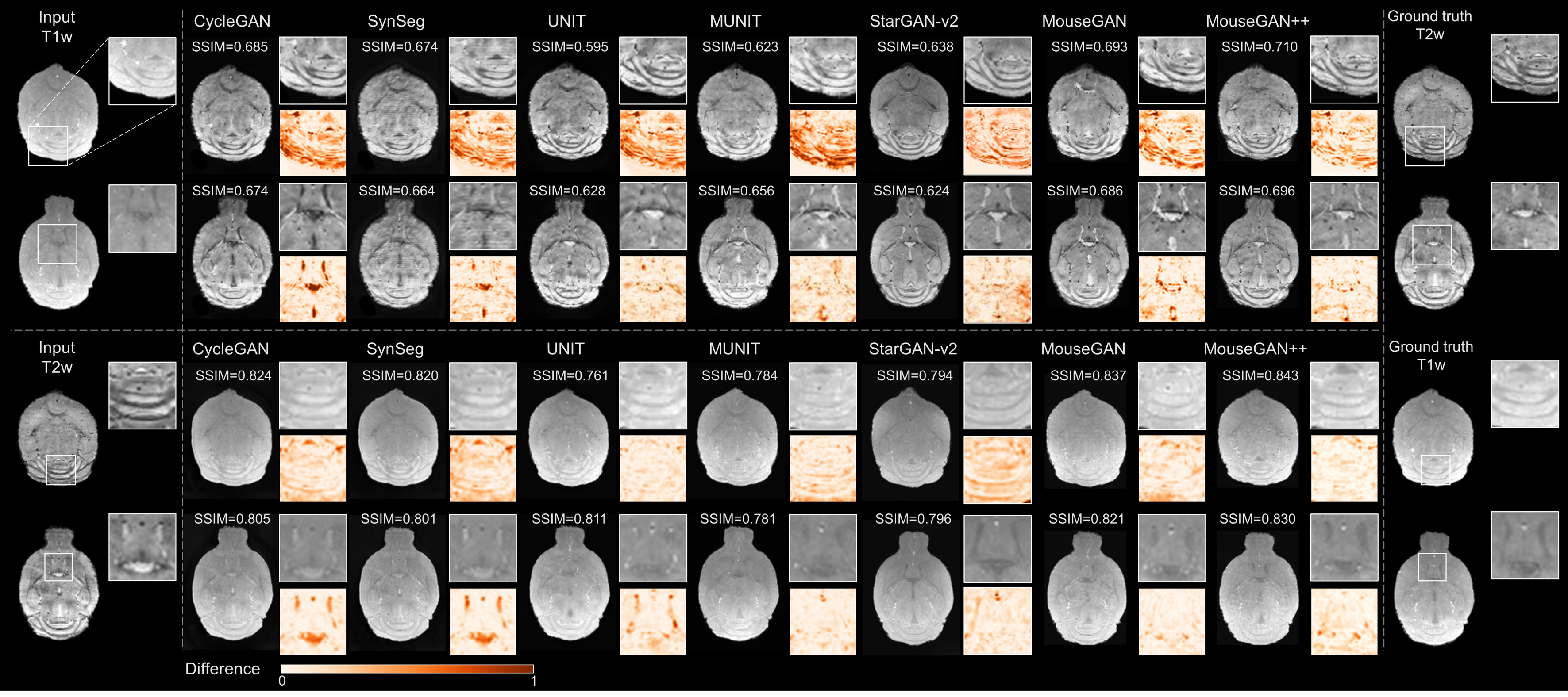}

\caption{Visualization of modality translation results between T1w and T2w. From left to right: raw input images (1st column), the synthetic images produced by different translation methods (2nd-7th column), and the ground truth target images (last column). For each real or synthetic image, we also show the zoomed-in regions of interest (ROI) in gray tunes as well as the zoomed-in residual image between the synthetic image and the ground truth of the same ROI. The colorbar indicates the magnitude of the normalised residual.}
\label{fig4}
\end{figure*}

\begin{figure}[hbt!]
\centering
\includegraphics[width=0.425\textwidth]{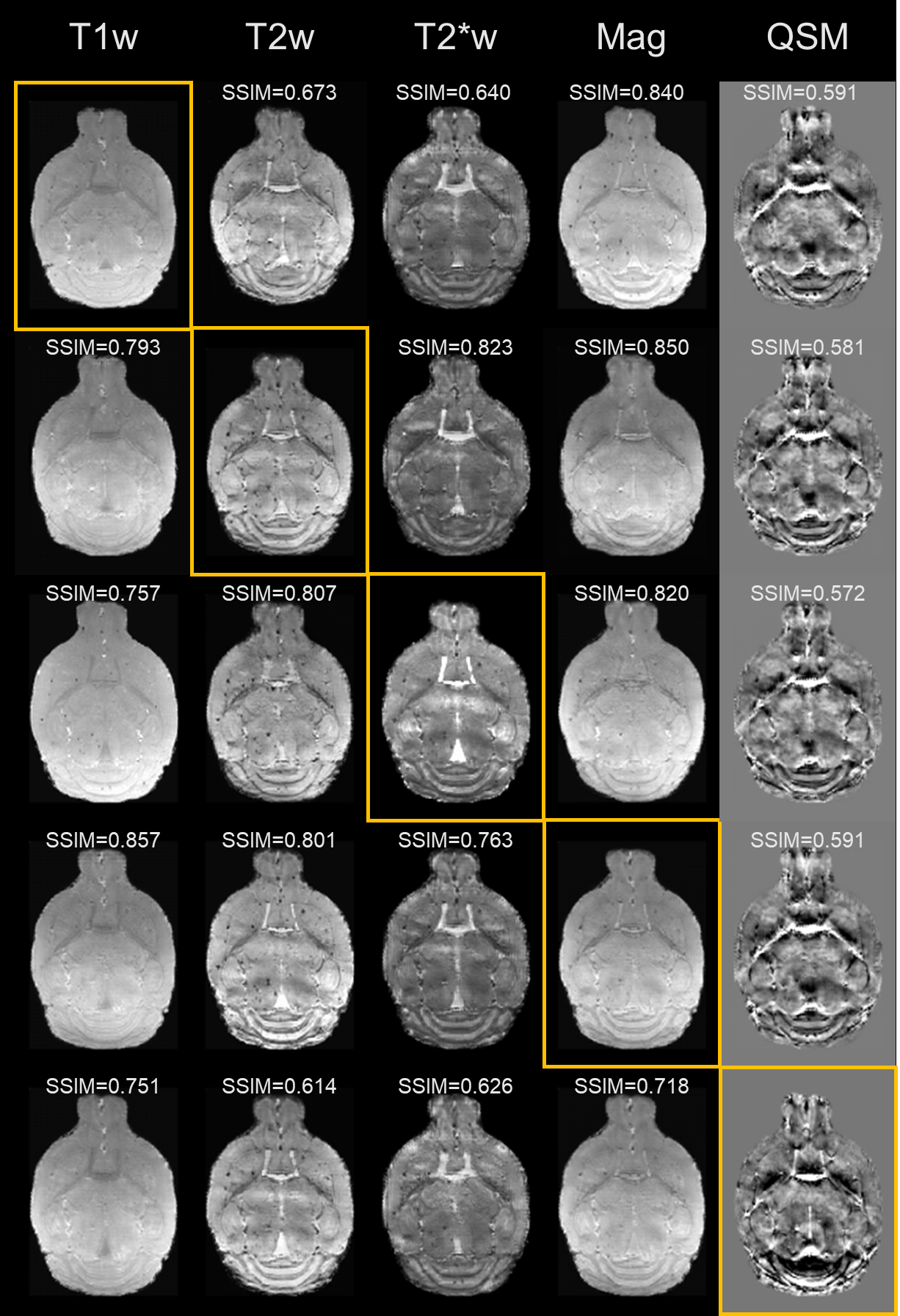}  

\caption{Visualization of MouseGAN++ modality translation results on all five modalities. From top to bottom are the exemplary samples from the results when T1w, T2w, T2*w, Mag, or QSM is used as a single input modality, respectively. Yellow boxes represent input real images for each row, and could also serve as ground truth for each column. SSIMs between the synthetic and ground-truth real images are shown on the top of each image.}
\end{figure}

\subsection{Evaluation on Two-Modality Image Translation}
As recent state-of-the-art methods mostly focus on one-to-one domain translation (i.e., one-to-one modality translation in our case), we first conducted T1w-T2w modality translation experiments in order to ensure comparability. The quantitative results of T1w-T2w translation are summarized in Table 1 and a few exemplary synthetic images are shown in Fig. 4. We can observe that for both two-direction translations, MouseGAN++ surpasses other methods in terms of all five metrics that are used to measure image synthesis quality. More specifically, for the translation from T1w to T2w, MouseGAN++ achieves the best performance with the highest SSIM of 0.716 and the lowest LPIPS of 0.173. We can also observe from Fig. 4 that MouseGAN++ produces sharper and more realistic textures, with the content better aligned with anatomy than other methods. For example, the lateral ventricles (zoomed-in ROIs from the second example) ought to have low contrast in T1w and, conversely, high contrast in T2w. Unlike MouseGAN++, synthetic images obtained from CycleGAN and SynSeg ignore these contrast features. While images synthesized by StarGAN-v2 show better contrast features, unexpected image deformation is present in the third ventricle and fourth ventricle (the second row, Fig.4). Moreover, MouseGAN++ yields significantly less pixel-level bias than others, especially in the cerebellar cortex. From our perspective, a fundamental reason for the better performance of MouseGAN++ is the attribute and content disentanglement, which enables more precise translation to different modalities. For the inverse translation task from T2w to T1w, MouseGAN++ achieves the best performance with an average SSIM of 0.841 and an average PSNR of 26.70, which is consistent with the T1w to T2w task. By contrast, the competing methods, CycleGAN, SynSeg, and StarGAN-v2, produce erroneous contrast or deformation in some brain regions during translation (zoomed-in ROIs in the last row).

\subsection{Evaluation on Multi-Modality Image Translation}

Here we further test MouseGAN++ on the five-modality translation task, where only images from a single modality were used as input to synthesize images of the other four modalities. Figure 5 reports the exemplary synthetic images along with the average SSIM for each test modality. Each row represents the test samples of the same subject from T1w, T2w, T2*w, Mag, and QSM, respectively.

We have two important observations from Fig. 5. First, some modalities are easier to synthesize than others: e.g., the averaged SSIMs of Mag and T1w being the target are 0.807 and 0.790, which are much higher than other modalities, e.g., QSM, with an averaged SSIM of 0.584. The suboptimal translation performance of QSM could be due to its immense different intensity distribution from other modalities, making this translation task particularly challenging. On the other hand, T2w, T2*w, and Mag can be good candidates for source modality, yielding average SSIMs of 0.762, 0.739, and 0.753 when being transferred into other modalities. By contrast, T1w and QSM generate lower SSIM (0.686 and 0.677, respectively). Particularly, we observe an asymmetric translation between T1w and T2w, the two most commonly used MRI modalities in practice: T2w $ \rightarrow $ T1w has a SSIM of 0.793, whilst T1w $ \rightarrow$ T2w lies only 0.673, which suggests that T2w could provide more structural information than T1w for mouse studies.

\begin{table*}[h]
\centering
\caption{Comparison of segmentation performance of different methods on T1w and T2w using multi-modality dataset}
\includegraphics[width=0.98\textwidth]{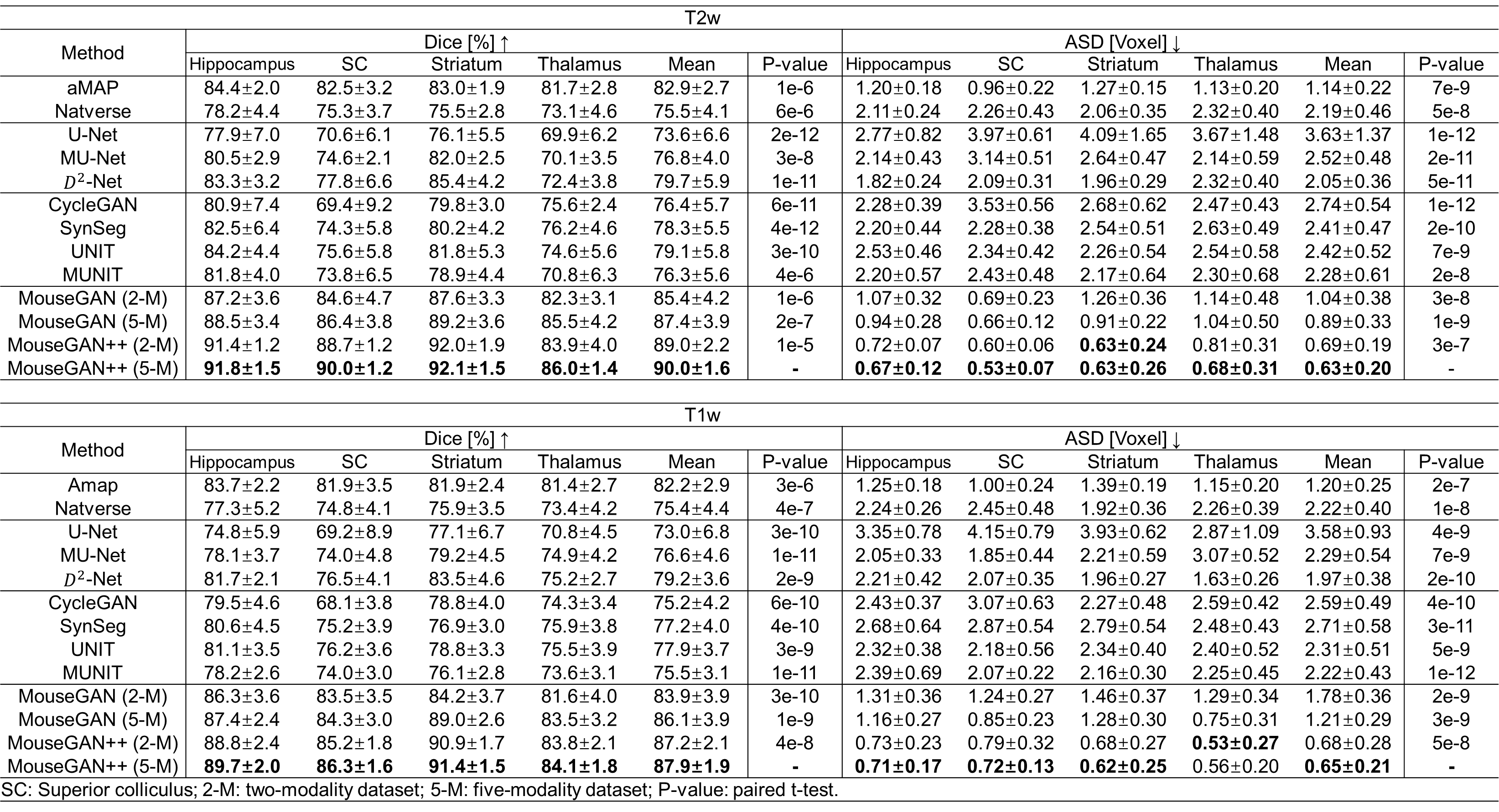}
\end{table*}

\begin{figure*}
\centering
\includegraphics[width=0.98\textwidth]{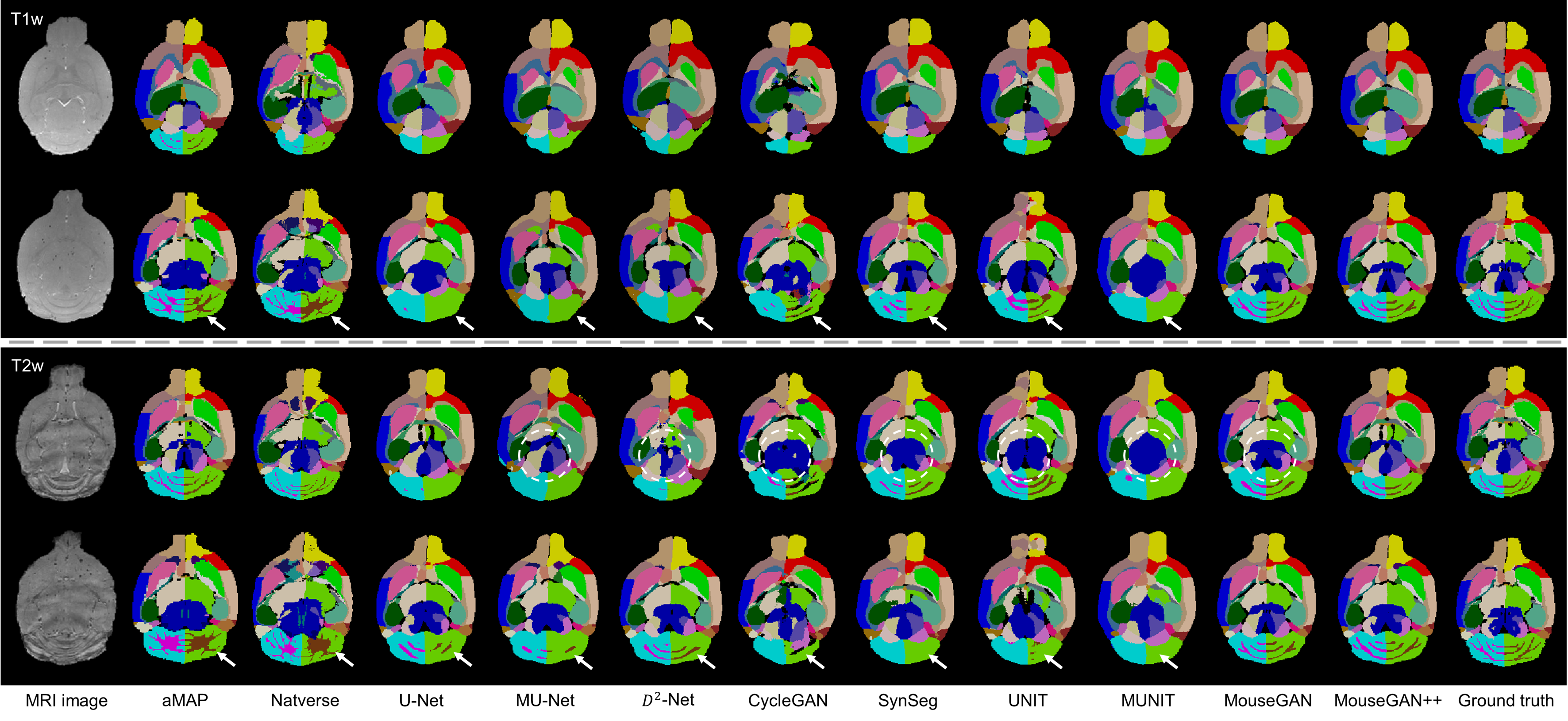}

\caption{Exemplary segmentation results on T1w and T2w. From left to right: images to be segmented(1st column), segmentation results of different comparative methods (2rd-11th column), results of our MouseGAN++ (12th column), and the ground truth segmentation (the last column). Our model can segment up to 50 mouse brain structures shown in different colors. White circles and arrows indicate erroneous segmentation details.}
\label{figX}
\end{figure*}

\begin{figure*}[h]
\centering
\includegraphics[width=0.95\textwidth]{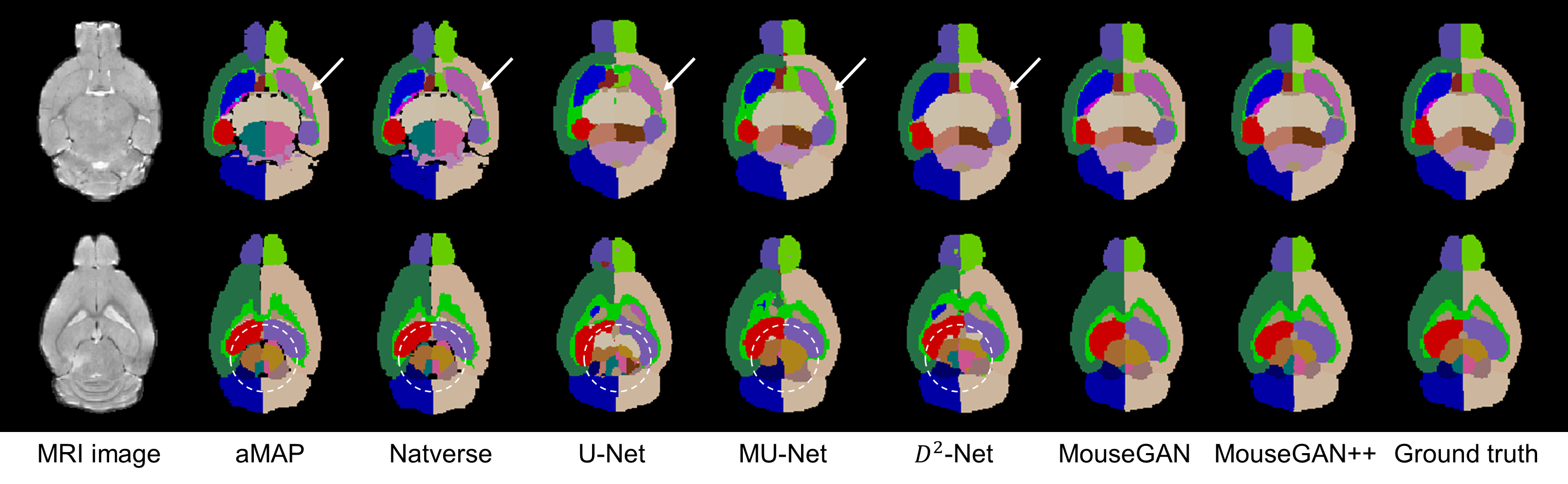}

\caption{Exemplary segmentation results on MRM NeAt T2w dataset. From left to right: images to be segmented(1st column), segmentation results of different comparative methods (2rd-7th column), results of our MouseGAN++ (8th column), and the ground truth segmentation (the last column). White circles and arrows indicate erroneous segmentation details.}
\label{fig7}
\end{figure*}

\begin{table*}[h]
\centering
\caption{Comparison of segmentation performance of different methods on T2w using MRM NeAt dataset}
\includegraphics[width=0.95\textwidth]{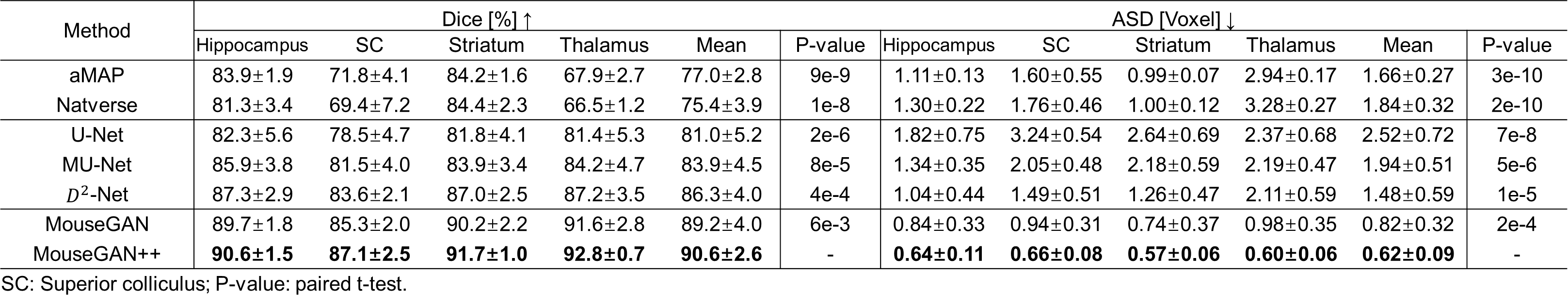}
\end{table*}

\subsection{Evaluation on Brain Structure Segmentation}

The quantitative results of brain structural segmentation on T1w and T2w, the two most important MR modalities used in mouse brain studies, are summarized in Table 2, with a few exemplary segmentation results presented in Fig. 6. In both modalities, MouseGAN++ consistently surpasses comparison methods, demonstrating the effectiveness of our proposed framework. Compared to the baseline U-Net segmentation with an average Dice of 73.6\%, conventional translation methods (CycleGAN, SynSeg, UNIT, MUNIT) contribute to an improved Dice of around 76.3\%-79.7\%, which could be attributed to the synthetic images that increase both the amount and the diversity of training data. As one of the SOTA disentanglement methods, $D^2$-Net also outperforms the baseline U-Net (yet inferior to MouseGAN and MouseGAN++), owing to the advantage of untangling during multi-modality training. In comparison, the two atlas-based segmentation methods, aMAP and Natverse, obtain average Dice scores of 82.9\% and 75.5\%, respectively. Their segmentation performance is largely determined by the quality of atlas registration. As we know, registration is not an easy task, as the heterogeneous contrasts existing in different image spaces might not provide enough guidance for intensity-based registration metrics, such as mutual information. Finally, MouseGAN++ achieves the best Dice scores (89.0\%) and generates delicate segmentation results, e.g., in the regions of deep brain structures and cerebellum (Fig. 6), which demonstrates that the content encoder in the modality transfer model can successfully encode the structural information shared between different modalities, thus mitigating the influence of the domain gap and the restriction of scarce information in a single modality. The additional incremented performance of MouseGAN++ as compared to MouseGAN could be due to a better disentanglement of the latent properties thanks to contrastive learning. Also, when making use of five modalities, MouseGAN++ gains a further 1\% increment of Dice and 0.06 lower distance of ASD compared to the two-modality version, demonstrating different modalities do provide complementary information. Last but not least, the segmentation performance of T2w is slightly better than T1w images, which may be due to the fact that T1w images provide less structural information than T2w, consistent with our previous modality transfer analysis in Section 4.B.

\begin{figure}[ht]
\centering
\includegraphics[width=0.45\textwidth]{figures/Fig8.png}

\caption{Comparison of translation training loss curves in ablation study between previous MouseGAN (blue) and MouseGAN++ (orange).}
\end{figure}

\begin{figure}[ht]
\centering
\includegraphics[width=0.45\textwidth]{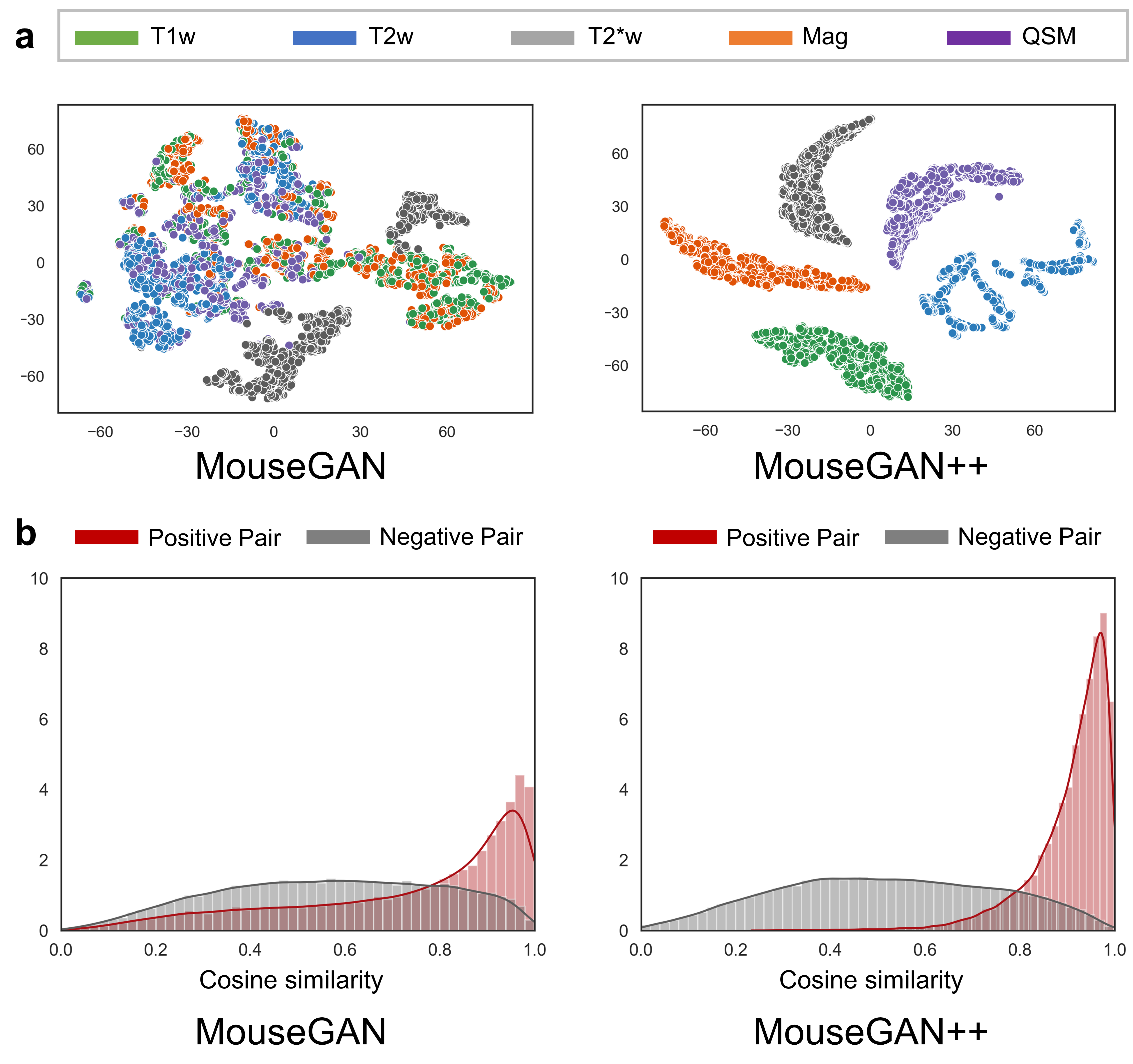}

\caption{Visualization of latent attribute and content embedding space without (MouseGAN) and with contrastive learning losses (MouseGAN++). (a) Scatter plot of attribute feature clusters using t-SNE. (b) The cosine similarity distributions of content features.}
\end{figure}

\subsection{Evaluation Over MRM NeAt Dataset}
It is interesting to compare the performance of MouseGAN++ with SOTA methods on an additional dataset, MRM NeAt. As shown in Table 3 and Fig. 7, atlas-based methods produce unsatisfied results as the registration procedure causes brain structural deformation, especially for the elongated structures (e.g., corpus callosum). Note that these small regions are also challenging for deep learning methods, especially when the training set is limited in size. Encouragingly, on this challenging small dataset, MouseGAN++ achieves the best Dice and ASD, outperforming U-Net, MU-Net, $D^2$-Net and the previous MouseGAN. Even without the translation module, the pretrained weight learnt from disentanglement representation in MouseGAN++ is superior to other networks due to the decoupled modality-agnostic knowledge, hence improving the performance of downstream tasks.

\begin{table}[h]
\centering
\caption{The ablation study for translation results taking T2w or T1w as a single input, respectively. The average metrics are based on the test modalities}
\includegraphics[width=0.48\textwidth]{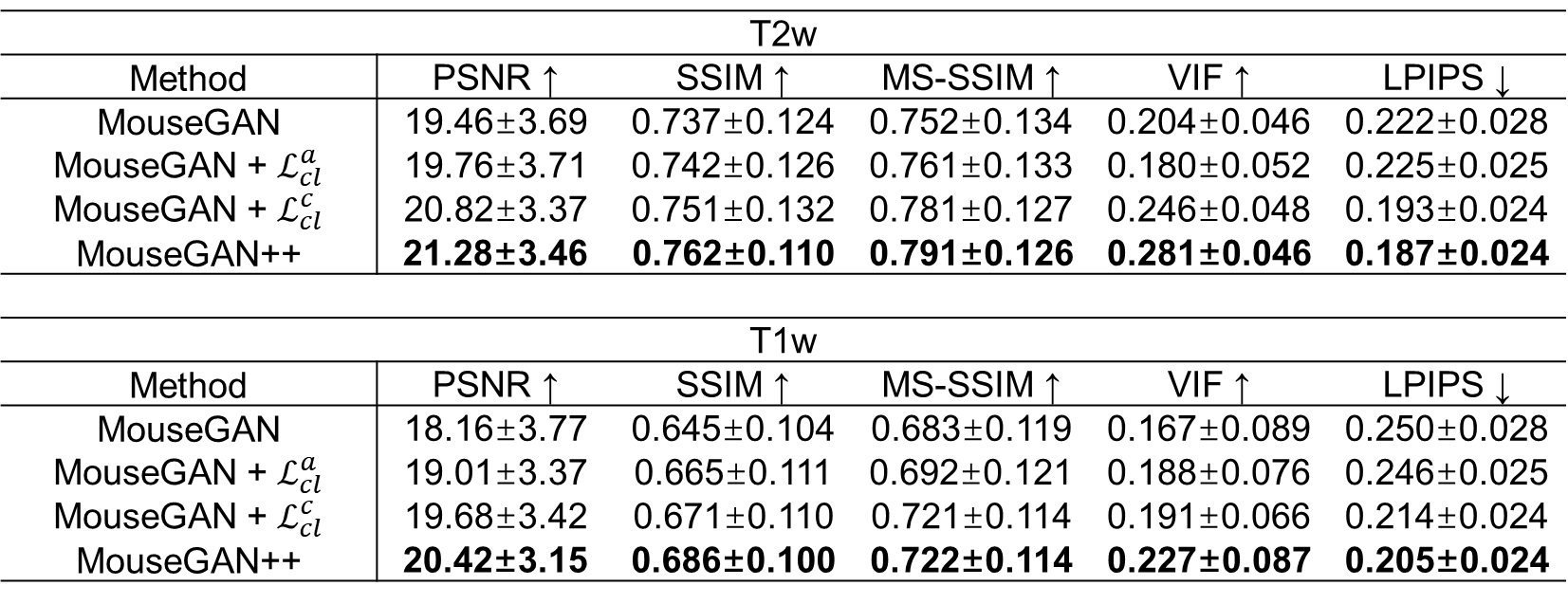}
\end{table}

\begin{table}[h]
\centering
\caption{The ablation study for exemplary brain structure segmentation evaluated by Dice scores on the 5-modality dataset}
\includegraphics[width=0.48\textwidth]{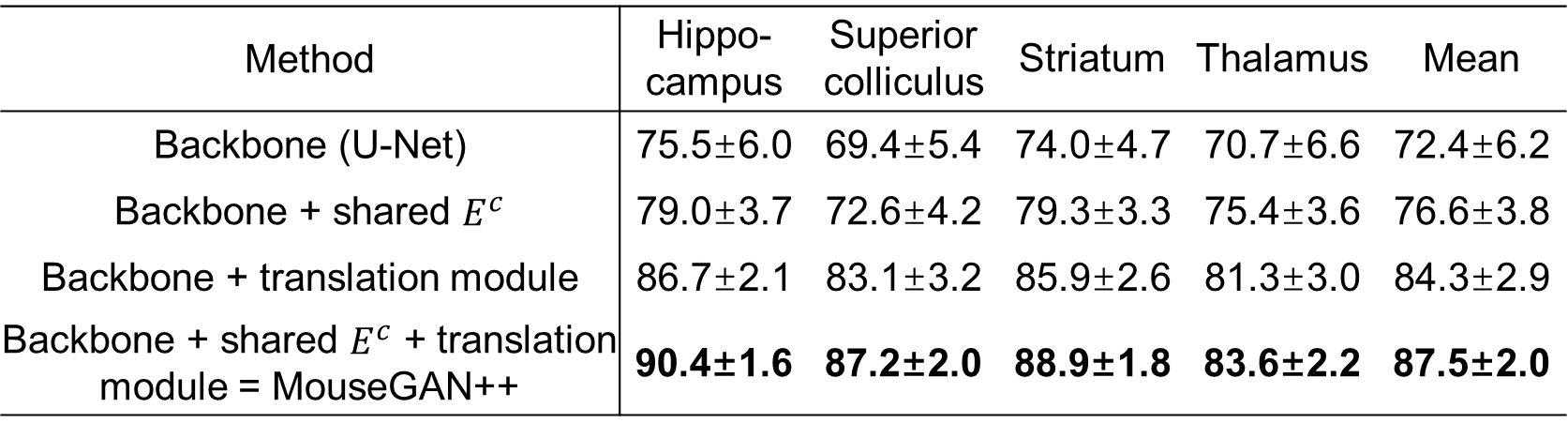}
\end{table}

\subsection{Ablation Study and Understanding of Contrastive Learning}
In this section, we investigate why MouseGAN++ further improves the performance of our previous MouseGAN, i.e., how attribute and content contrastive learning improves modality translation. We first conduct an ablation experiment on the five modality translation tasks to show the effectiveness of each contrastive learning strategy. The baseline method is MouseGAN, which is deployed without attribute $ {\cal L}_{cl}^a$ and content $ {\cal L}_{cl}^c$ contrastive losses. To fully understand the advantages of our method, we exhibit the following experiment results and analyses from three aspects, the effectiveness of contrastive losses over mouse brain structural MR images, visualization of latent spaces, and comparison between Gaussian prior and contrastive learning prior.

\subsubsection{Effectiveness of Contrastive Losses}
As shown in Table 4, compared with the baseline, the attribute contrastive loss alone gains an average SSIM of 0.665 and 0.742 when taking T1w and T2w as input, respectively. A similar performance increase is observed when we add the content contrastive loss alone (0.671 to 0.751). Adding both components together makes MouseGAN++, which achieves 0.686 and 0.762 in terms of average SSIM. This incessant increase illustrates that attribute and content discrimination can be jointly implemented to produce better disentanglement representation in translation tasks. In Fig. 8, we present the training loss curve of self-modality reconstruction $ {\cal L}_1^{self-recon}$ as well as $ {\cal L}_1^{cyc}$, which is associated with the quality of translation. The MouseGAN++ with contrastive losses shows a more stable training procedure than the original MouseGAN as the adversarial losses achieve the balance between the discriminator and generator in the multi-modality dataset.

One corner case we intend to clarify here is how the network knows to deal with two very similar mouse brains in one mini-batch. Even though this case can be deemed as a hard example for network discrimination, the anatomical contents from the same subject should be located closest to each other in the latent content space despite the presence of modality translation. Additionally, some works report that attribute features may be further divided into domain-specific style and domain-invariant style \cite{liu2021semi-R3Q5, ilse2020diva-R3Q5}. In MouseGAN++, the modality-related (domain-specific) features are preserved in attribute space, while the modality-invariant (domain-invariant, or in other words, structure-related) information is condensed in the content space, which is adequate for MR image disentanglement.

Furthermore, we qualitatively analyze the contribution of each individual module of MouseGAN++ with regard to segmentation. We first set up the backbone of the network as a baseline model without shared content encoder weight and translation module (baseline U-Net). Then we add each module one by one into the baseline model. We evaluate different methods on five modalities with every single modality as input successively and compute the average metrics. As shown in Table 5, both the shared content encoder and the translation module bring significant performance advancement, indicating that each module can work collaboratively in our proposed framework.

\subsubsection{Visualization of Latent Spaces}
To illustrate how contrastive learning improves attribute and content feature embedding themselves, we visualize attribute features of five modalities via t-SNE \cite{van2008tSNE}. As shown in Fig. 9a, the attribute features of our proposed methods are well separated in attribute space, while the baseline method confuses all modality features except T2*w modality, which further demonstrates the capacity of MouseGAN++ to achieve better attribute discrimination.

Then we further evaluate the instance-specific content discrimination. We calculate the cosine similarity between content features from the input modality and the modality-translated images of the same subject (positive pairs), as well as features from different subjects (negative pairs). As shown in Fig. 9b, a better feature embedding is denoted by a more separable distribution between positive and negative samples, which demonstrates that MouseGAN++ is able to learn more instance-specific and less modality-relevant features, hence conducting better semantic information retention during modality translation.

\subsubsection{Gaussian Prior vs. Contrastive Learning Prior}
By contrast with existing approaches \cite{wagner2021structure-R4Q2-2, havaei2021conditional-R4Q2-3, wang2022sgdr-R4Q2-4, bercea2021FedDis-R4Q2-5} that assume the Gaussian approximation over the latent attribute space as the primary bias or prior to promote decoupling, we claim that the contrastive learning prior is more appropriate for approaching an ideal latent distribution of a multi-modality dataset. Unlike MouseGAN, which follows Gaussian priors, MouseGAN++ follows contrastive learning priors. The comprehensive results of the above-mentioned comparisons, ablation study, and visualization of latent space demonstrate that the MouseGAN++ achieves better performance as well as translation quality compared with SOTA methods.

Nowadays, in practice, multimodal images are commonly unpaired, which can be a major obstacle to deploying such paired training methods \cite{yang2022D2-Net}. Our efforts also extend this promising direction via fusing cross-modality information in an unpaired manner and yielding more robust performance when the modality is missing. Another point we want to emphasize is the maximum flexibility of our framework as we trained separately rather than jointly. This is reflected in the flexibility in choosing pretext tasks, which is important for disentanglement and downstream tasks. In order to impute a missing modality, we chose modality translation as our pretext task in this work. However, one limitation of our work is the dilemma caused by domain shifts, since the image quality and contrast from various centers may differ greatly. An appealing and promising solution is to convert our pretext task to cross-center image translation so that the learned center-agnostic features in the content space would alleviate the segmentation performance degradation.

\section{Conclusion}
In summary, we propose a novel GAN-based framework, MouseGAN++, for simultaneous image synthesis and segmentation for mouse brain MRI. Based on a disentangled representation of content and style attributes strengthened by contrastive learning, MouseGAN++ is able to synthesize multiple MR modalities from single ones in a structure-preserving manner and can hence handle cases with missing modalities. Furthermore, it uses the learned modality-invariant information to improve structural segmentation. Our results demonstrate that MouseGAN++ achieves significant performance improvement over both translation and segmentation tasks and has the potential to be applied in more neuroimaging applications.

The broader implications of our work are the publicly available packaged pipeline provided by MouseGAN++ and the multiple modality MR dataset for facilitating preclinical research, especially for the communities interested in rodent brain.
In addition, the unsupervised pretext task alleviates the cost of deployment and promotes the clinical use of cutting-edge machine learning techniques. In the future work, one of the attractive directions is to integrate our previous brain extraction tool \cite{yu2022ben} with MouseGAN++ as an end-to-end neuroimaging processing pipeline.

\bibliography{tmi} 
\bibliographystyle{unsrt}

\end{document}